\documentclass[12pt]{iopart}

\usepackage{graphicx}
\def\dbar{{\mathchar'26\mkern-12mu d}}
\usepackage{color}

\newcommand{\DyT}{Dy$_{2}$Ti$_{2}$O$_{7}$}

\newcommand{\HoT}{Ho$_{2}$Ti$_{2}$O$_{7}$}
\begin{document}

\title[]{Determination of the Entropy via Measurement of the Magnetization: Application to the Spin ice \DyT}

\author{L. Bovo$^1$ and S. T. Bramwell$^1$}

\address{1. London Centre for Nanotechnology and Department of Physics and Astronomy, University College London, 17-19 Gordon Street, London WC1H 0AJ, UK.}

\ead{l.bovo@ucl.ac.uk}

\begin{abstract}
The residual entropy of spin ice and other frustrated magnets is a property of considerable interest, yet the usual way of determining it, by integrating the heat capacity, is generally ambiguous. Here we note that a straightforward alternative method based on Maxwell's thermodynamic relations can yield the residual entropy on an absolute scale. The method utilises magnetization measurements only and hence is a useful alternative to calorimetry. We confirm that it works for spin ice, \DyT, which recommends its application to other systems. 
The analysis described here also gives an insight into the dependence of entropy on magnetic moment, which plays an important role in the theory of magnetic monopoles in spin ice. Finally, we present evidence of a field-induced crossover from correlated spin ice behaviour to ordinary paramagnetic behaviour with increasing applied field, as signalled by a change in the effective Curie constant. 

\end{abstract}

\maketitle

Residual entropy is an essential feature of highly frustrated spin models, as it reflects the macroscopic ground state degeneracy inherent to such systems~\cite{Diep}. 
Experimental realisations of frustrated spin models are numerous, for example, see Refs.~\cite{RamirezEspinosa,LeeBroholm,Mireb}. In many of these systems the degeneracy is removed by ordering perturbations, but in spin ice~\cite{Harris, Ramirez}, and some other rare earth magnets (see for example Ref. ~\cite{Novikov}), the entropy is observed to reach a finite low temperature limit on experimental time scales.  

The usual way of determining magnetic entropy starts by determining the entropy increment by integration of the experimental heat capacity:
\begin{equation}\label{one}
\Delta S \equiv S(T)-S(T_0) = \int_{T_0}^T \frac{C_0}{T}dT,
\end{equation}
where $C_0$ is the magnetic contribution to the heat capacity at zero applied field. In order to convert the estimated $\Delta S$ into an absolute entropy $S(T)$ it is necessary to know the entropy at a particular temperature $T$. In the case of a spin ice~\cite{Harris} such as \DyT~\cite{Ramirez} it is expected that $S(T \approx 10 ~{\rm K})$ is approximately $2nR\ln(2)$ on account of the thermal population of only one crystal field doublet per Dy (here $n$ is the molar amount of  \DyT). In this way it has been inferred that as $T_0\rightarrow 0$ the entropy approaches not the third law value, $S=0$, but instead the Pauling value $S= n R \ln(3/2)$ characteristic of a degenerate ground state controlled by `ice rules'.  A recent work \cite{Kycia} found corrections to the Pauling entropy, but these are only measurable on timescales much longer than those of normal calorimetry experiments, and may be neglected for our purposes.

The residual entropy gives the bluntest possible measure of a disordered magnetic state, but one that gives basic and valuable insight to the nature of that state, in complement to more detailed measures such as neutron scattering~\cite{Harris,Fennell,Yavorskii}. Presuming that it is possible to isolate $C_0$, the weakest part of the calorimetric analysis is generally the difficulty in estimating the absolute entropy at a given temperature. 
One can contrast the latter with the case of molecular systems like water ice~\cite{Giauque} which can be driven into the gas phase to allow entropies to be accurately estimated on the basis of spectroscopic parameters. In magnetism, although one might achieve something similar by heating to the paramagnetic phase, the practical difficulty of accurately estimating both the paramagnetic entropy and the correction arising from lattice vibrations, are generally overwhelming. Spin ice represents a fortunate exception, in which the lattice contribution is weak at 
$T \approx10$ ${\rm K}$ and the paramagnetic contribution arises from single crystal field doublet, well separated from other states. In the case of most other magnetic systems, such a fortunate coincidence is not available. An example is ${\rm Tb_2Ti_2O_7}$, another frustrated magnet, closely related to spin ice. In that case the entropy is a quantity of particular importance in distinguishing microscopic models~\cite{Mirebeau}, but it cannot be easily put on an absolute scale.   
 
An alternative way of estimating the entropy exploits exact thermodynamic relationships. In magnetic thermodynamics the conjugate thermodynamic variables that define magnetic work are the magnetic moment $I$ and the internal H-field $H_{\rm int} = H_{\rm applied}- \mathcal{D} M$ where $M=I/V$ is the magnetization, $V$ is the volume and $\mathcal{D}$ is the demagnetizing factor (assuming an ellipsoidal sample). The incremental magnetic work of reversible magnetization is $\dbar W = \mu_0 H_{\rm int}dI $.  
The magnetic moment is related to the entropy by the Maxwell relation: 
\begin{equation}\label{main}
\left(\frac{\partial I}{\partial T}\right)_{H_{\rm int}} = \frac{1}{\mu_0}\left(\frac{\partial S}{\partial H_{\rm int}}\right)_T.
\end{equation}
Integration of this equation gives:  
\begin{equation}\label{integral}
\mu_0 \int_0^{H_{\rm int}} \left(\frac{\partial I (T,H_{\rm int})}{\partial T}\right)_{H_{\rm int}} d H_{\rm int} = 
S(T,H_{\rm int}) - S(T,0). 
\end{equation}
Thus, if the maximum applied field is strong enough to remove all magnetic entropy, such that $S(T,H_{\rm int} = H_{\rm max})= 0$, it is then possible to estimate $S(T) \equiv S(T,H_{\rm int} = 0)$ on an absolute scale at any temperature. This  requires that any other field-induced thermodynamic changes (for example magnetostriction) are negligible, a safe assumption for most substances.
 
While the above method of entropy determination is exact in theory, one would be ill advised to accept its results in the absence of a control experiment, for in practice, experimental uncertainties could easily compound to undermine the measurement principle. The purpose of present note is to report such a control experiment, which can be used as a reference point for other studies. Spin ice lends itself well to this experiment as the magnetic entropy is uncontroversial, having been confirmed by many authors (e.g. Refs.~\cite{Ramirez,Higa,Kycia}). It should be noted that Aoki {\it et al.}~\cite{Aoki} studied the entropy of spin ice in the millikelvin range via the magnetocaloric effect: a related, but more specialised method to the one discussed here. 

Experimentally, we measured the magnetization at different temperatures as a function of applied field on two different systems, a Quantum Design SQUID magnetometer ($\mu_0 H_{\rm max} = 7$ ${\rm T}$) and a Vibrating Sample Magnetometer (VSM) measurement system for the Quantum Design PPMS ($\mu_0 H_{\rm max} = 14$ ${\rm T}$). Magnetic fields were applied along the $[111]$ axis of a $1.42 \times1.42 \times 0.57$ mm$^3$ crystal of \DyT~(cubic, space group Fd-3m), [111] being parallel to its shortest dimension. The same crystal was used to measure the specific heat by means of a Quantum Design PPMS equipped with 3He-Probe to measure down to $0.4$ ${\rm K}$ in order to estimate the magnetic entropy via the standard~\cite{Ramirez} calorimetric method of Eqn. {1}.  

The magnetic moment $I$ ${\rm vs.}$ $H_{\rm applied}$ was measured at many temperatures in the range $1.8 \le T / {\rm K} \le 10$ (typically every $0.1$ ${\rm K}$). The applied field was corrected for the demagnetizing field to give $H_{\rm int}$ in the standard way, although we used an experimentally determined demagnetizing factor in line with the result of Ref.~\cite{Laura}. Subsequently, experimental data were interpolated first, in order to extrapolate the magnetic moment as a function of a specific set of $H_{\rm int}$, at all given temperatures. In this way, the magnetic moment $I$ ${\rm vs.}$ $H_{\rm int}$ and its temperature derivative $-(\partial I/\partial T)_{H_{\rm int}}$ was calculated at constant internal field $H_{\rm int}$, up to the maximum applied field. Fig. 1 shows the latter quantity as a function of field for selected temperatures. For each temperature, an optimised temperature step $\Delta T$ was determined in order to minimise spurious effects and make an unbiased estimate of $I (T,H_{\rm int})~{\rm vs.}~T$. In order to do so, for each temperature, the gradient $\Delta I/\Delta T$ was calculated around the centred value $T_0$ and $T_0 \pm 0.1$ ${\rm K}$, for three different values of $\Delta T$ corresponding to the forward and reverse finite increments $\Delta T = \pm \delta$ and the centred increment $\Delta T = 2 \delta$. The parameter $\delta$ was then chosen to be as large as possible under the constraint that the nine different estimates of the derivative $(\approx \Delta I/\Delta T)_{H_{\rm int}}$ tended to be equal. Their absolute minima and maxima fluctuations were taken as error bars. Typically, we found $\delta = 1$ ${\rm K}$ at high temperature above $6$ ${\rm K}$, $\delta = 0.5$ ${\rm K}$ for intermediate temperatures and $\delta = 0.1$ ${\rm K}$ below $3$ ${\rm K}$. 

The data of Fig. 1 was transformed into the entropy difference $[S(T, H_{\rm int} = 0) - S(T, H_{\rm int})]$ using Eqn.s \ref{main} and \ref{integral}, by integrating the estimated $-(\partial I/\partial T)_{H_{\rm int}} \approx - {\left(\Delta I/\Delta T\right)}_{H_{\rm int}}$ with respect to the internal field. In Fig. 2 we show the result for the entropy per mole Dy, $s(T,H_{\rm int}) = S(T, H_{\rm int})/2n$. At low temperature, {$T \le 3$ {\rm K}}, the $s(H_{\rm int})$ curves show a distinct plateau that may be understood in terms of `kagome ice'~\cite{Mats-kag,Aoki}, where 1/4 of the Dy magnetic moments are pinned by the applied field. 

In Fig. 3 we compare $s(T)\equiv s(T, H_{\rm int} = 0)$ derived by the magnetometry method with that derived by the calorimetric method of Eqn. \ref{one}. The calorimetric entropy was calculated from the experimental specific heat measurement using the standard procedure of Ref.~\cite{Ramirez}. Here the scale of the calorimetric entropy was fixed by shifting the experimental data by $s(T=0)=1.686$ ${\rm JK^{-1}mol_{\rm Dy}^{-1}\approx (1/2) R \ln(3/2)mol_{\rm Dy}^{-1}}$. 

Referring to Fig. 3, the magnetometry method gives results that are in close agreement with the calorimetric method, albeit with larger error bars, and with some small systematic deviations evident at high temperature. A combination of the methods determines the residual entropy on an absolute scale without making any assumption of the high temperature entropy. Of course, there is no surprise that thermodynamics is obeyed, but our result does confirm that the magnetometry method can be used as a practical means of determining the residual entropy in situations where the calorimetric method is inconvenient or ambiguous. 

Finally it is interesting to consider our data in the context of emergent magnetic monopoles in spin ice~\cite{CMS,Ryzhkin}. The entropy as a function of magnetic moment plays an important role in the non-equilbrium thermodynamic approach to the motion of these magnetic charges~\cite{Ryzhkin}. There, as in the Jaccard theory of water ice, the entropy may be  assumed to depend on a configuration vector $\vec{\Omega}$ that is simply the magnetization divided by the monopole charge: $\vec{\Omega} = {\bf M}/Q$~\cite{Ryzhkin}. The field dependence of the molar entropy is given by:
\begin{equation}\label{m2}
s(M)-s(0) = \frac{\mu_0V_m|{\bf M}|^2}{2\chi_{\rm T} T },
\end{equation}
where $\chi_{\rm T}$ is the isothermal susceptibility and $V_m$ is the molar volume. 
As the quantity $\chi_{\rm T} T$ is approximately independent of temperature in the temperature range considered~\cite{Laura}, the entropy should be a linear function of the square of the magnetization or magnetic moment, that is nearly temperature-independent. This is born out by our data, shown in Fig. 4, where the expression \ref{m2} is confirmed, and the limit of the quadratic dependence is clearly visible. 

Fig. 4 raises an interesting point concerning the susceptibility of spin ice. In recent work it has been shown that the quantity $\chi_{T} T/C$ (where $C$ is the Curie constant) gradually 
rises above unity below about $T = 20 ~{\rm K}$, on account of long range correlations in the spin ice state~\cite{Jaubert-TSF}: a careful test of this is given in Ref. \cite{Laura}. If one defines 
$\chi_{\rm T}({\rm experimental}) = \mathcal{C}(T)/T$ then we can test that the experimentally derived $\mathcal{C}(T) > C$ imparts the correct slope to the graph of entropy versus magnetization squared in the limit of small magnetization. As shown in the inset of Fig. 4 there is complete consistency in this regard, but it is noteworthy that there is a crossover at small finite $M$ to a regime of larger slope where the experimental data is described by $\chi_T=C/T$ (red line in Fig. 4, main figure). This suggests that the long range correlations that cause $\mathcal{C}$ to be greater than $C$ are suppressed by a relatively small applied field, such that spin ice behaves as an ordinary paramagnet.  It would be interesting to extend this analysis to lower temperature, $T < 1$ ${\rm K}$, where the theory of Ref.~\cite{Ryzhkin} is more directly applicable. 

Another angle on this result may be gained by defining an effective susceptibility $\chi_{\rm eff} = M/H$ and plotting this versus field (Fig, 5). As expected, there is a close agreement with the expected susceptibility in the low field limit (inset, Fig. 5), but at higher fields there is a gradual departure of $\chi_{\rm eff}$ from the zero field value.  These results show that considerable care must be taken when measuring the susceptibility of spin ice materials, as stressed in Ref. \cite{Laura}.

\ack
It is a pleasure to thank R. Thorogate for technical assistance and G. Aeppli, J. Bloxsom 
for related collaborations. This research was supported by EPSRC Grant EP/I034599/1.

\newpage
\begin{figure}
 \begin{center}
 \renewcommand{\figurename}{Fig.}
\includegraphics[width=0.7\linewidth]{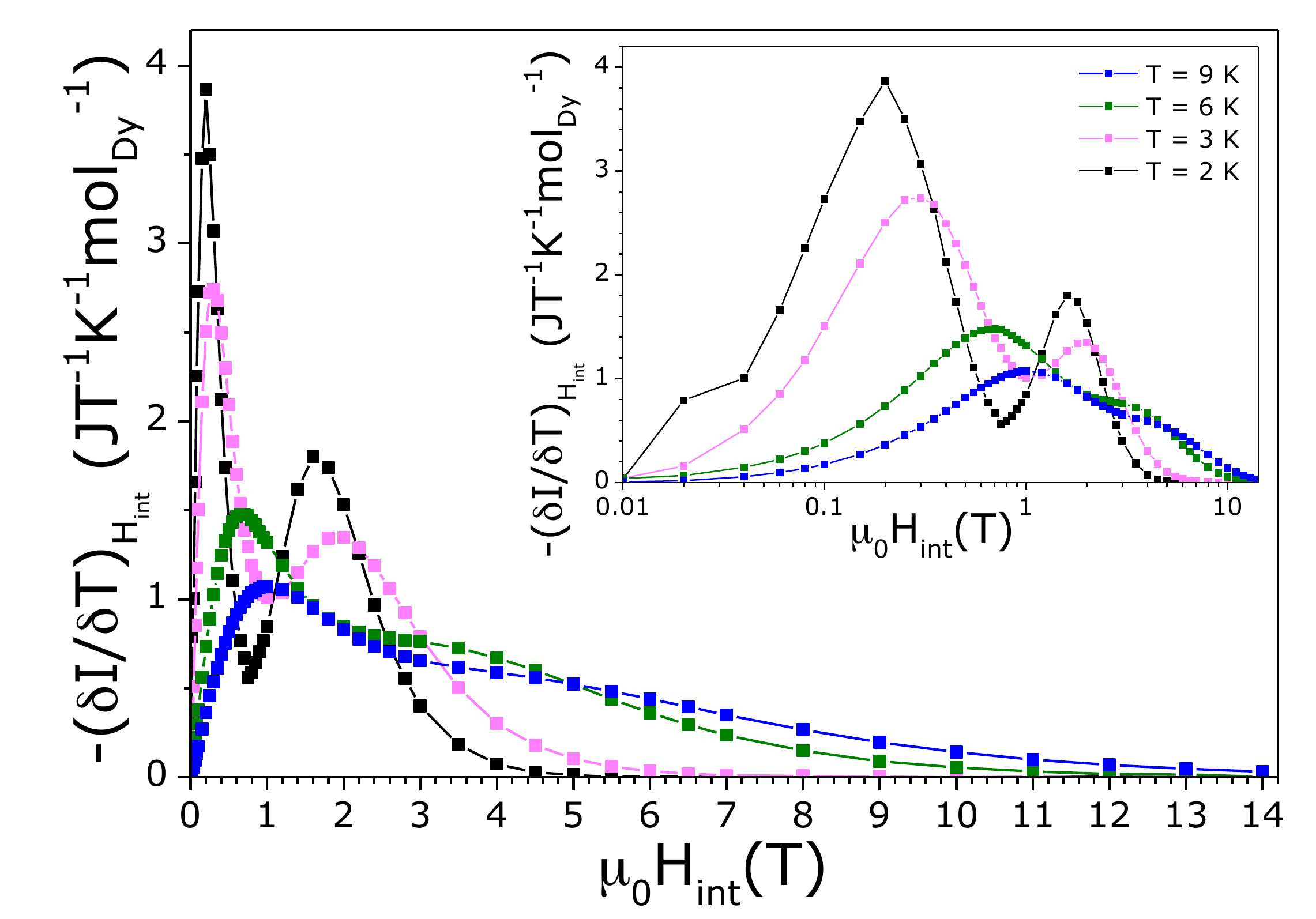}
\caption{Estimated temperature gradient of the magnetic moment, $-(\partial I/\partial T)_{H_{\rm int}}$ as a function of the internal field, for selected temperatures. The inset shows the same plot in logarithmic scale. Lines are guides to the eyes.}
\end{center}
\end{figure}

\begin{figure}
 \begin{center}
 \renewcommand{\figurename}{Fig.}
\includegraphics[width=0.8\linewidth]{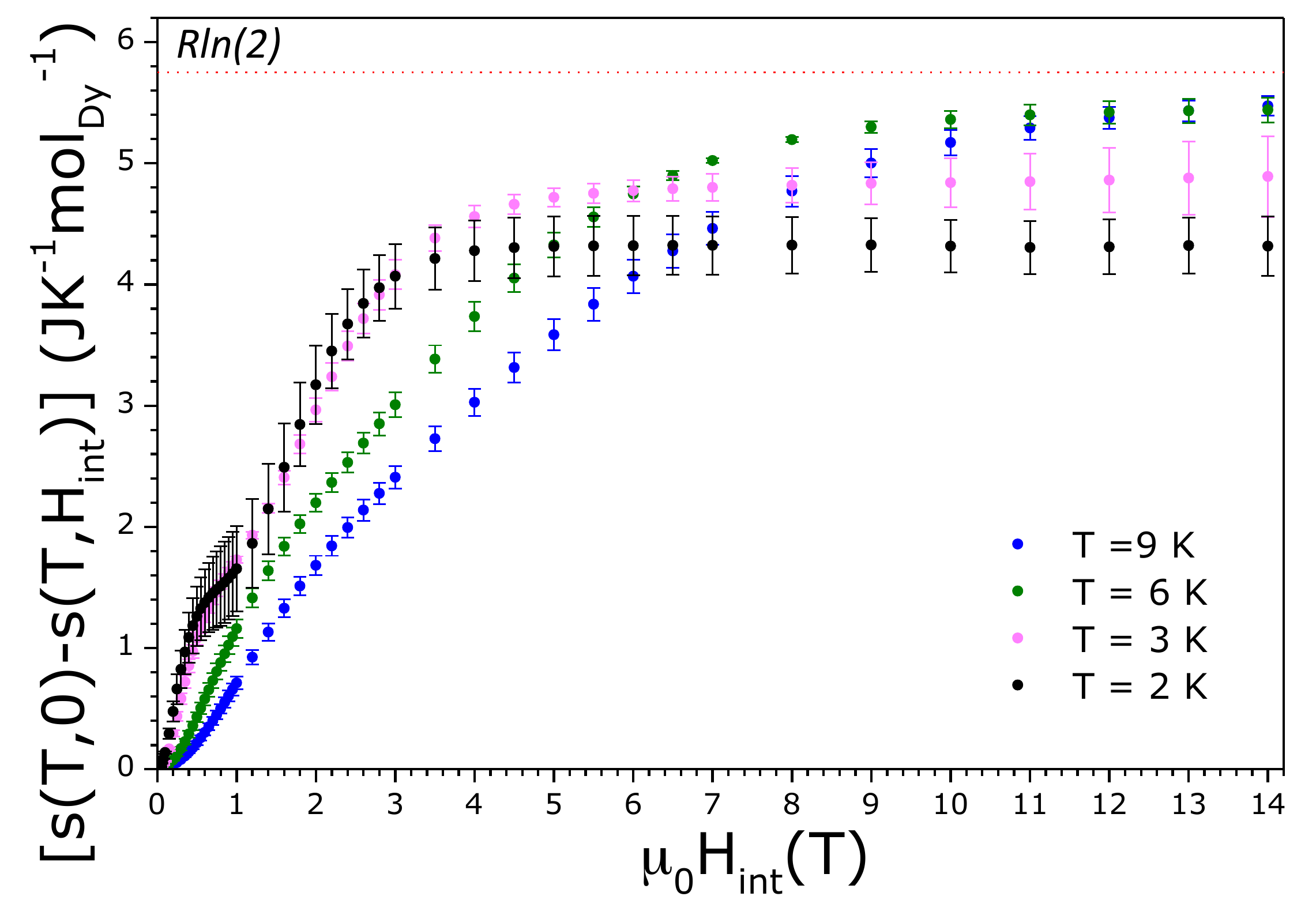}
\caption{Magnetic field dependence of $[s(T, H_{\rm int} = 0) - s(T, H_{\rm int})]$ for selected temperatures. Error bars represent absolute minima and maxima, see text for details. Red dotted line is the expected total entropy of the paramagnetic system ${\rm R\ln(2) mol_{\rm Dy}^{-1}}$ \cite{Harris, Ramirez}.}
\end{center}
\end{figure}

\begin{figure}
 \begin{center}
 \renewcommand{\figurename}{Fig.}
\includegraphics[width=\linewidth]{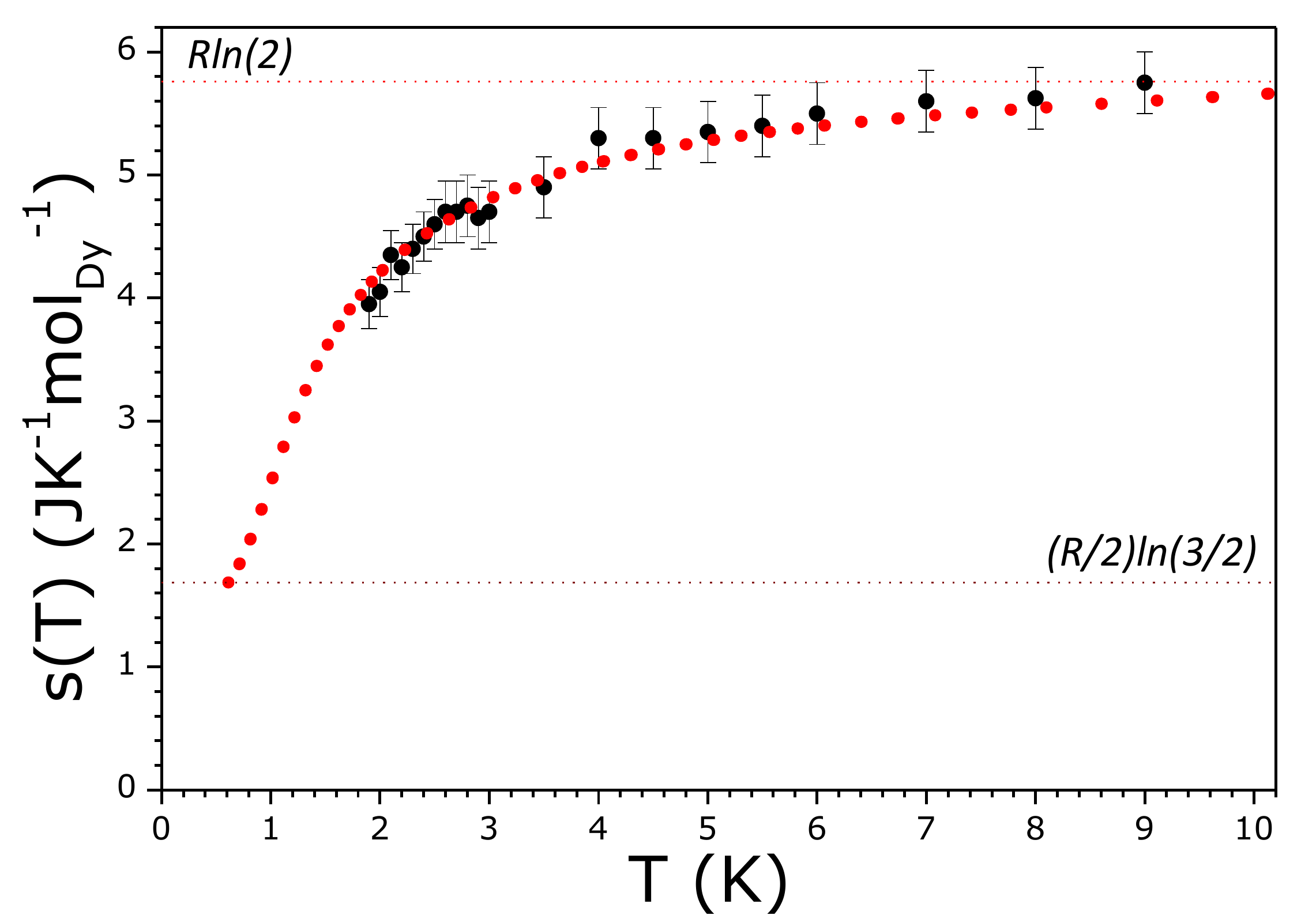}
\caption{Comparison between $s(T)\equiv s(T, H_{\rm int} = 0)$ derived by the magnetometry method (black dots) with that derived by the calorimetric method (red dots). Error bars represent absolute minima and maxima. Red dotted line is the expected total magnetic entropy of the system ${\rm R\ln(2)mol_{\rm Dy}^{-1}}$ \cite{Harris, Ramirez}. Purple dotted line represents the offset, $s(T=0)=1.686$ ${\rm JK^{-1}mol_{\rm Dy}^{-1}}$, by which the calorimetric data has to be shifted in order to be in absolute scale. The experimental offset agrees very well, with the expected Pauling's zero-point entropy \cite{Harris, Ramirez}.}
\end{center}
\end{figure}

\begin{figure}
 \begin{center}
 \renewcommand{\figurename}{Fig.}
\includegraphics[width=0.8\linewidth]{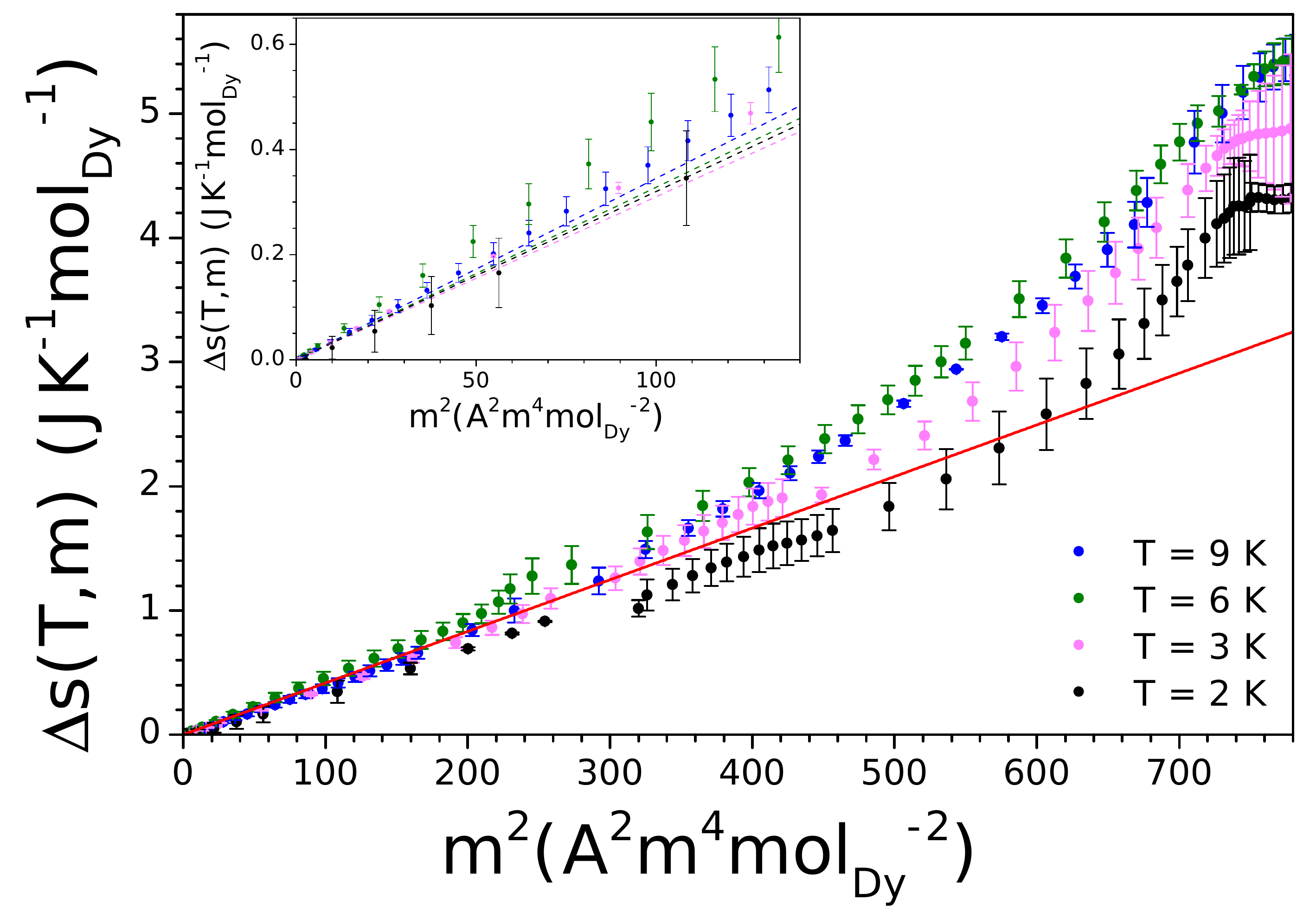}
\caption{Entropy increment $\Delta s(T)$ as a function of $m^2$ where $m$ is the molar magnetic moment. The red line is the expression derived from Eqn. \ref{m2} using $\chi_T = C/T$, where $C = 4.0$ {\rm K} is the paramagnetic Curie constant for \DyT. The inset shows the low field region where the dotted lines show the experimental amplitudes $\mathcal{C} > C$ defined by $\chi_T = \mathcal{C}/T$, as measured in Ref. \cite{Laura}. The same colour code is maintained throughout. The experimental data is better described using $\mathcal{C}$ in the low field limit, but using $C$ at stronger fields.}
\end{center}
\end{figure}

\begin{figure}
 \begin{center}
 \renewcommand{\figurename}{Fig.}
\includegraphics[width=0.8\linewidth]{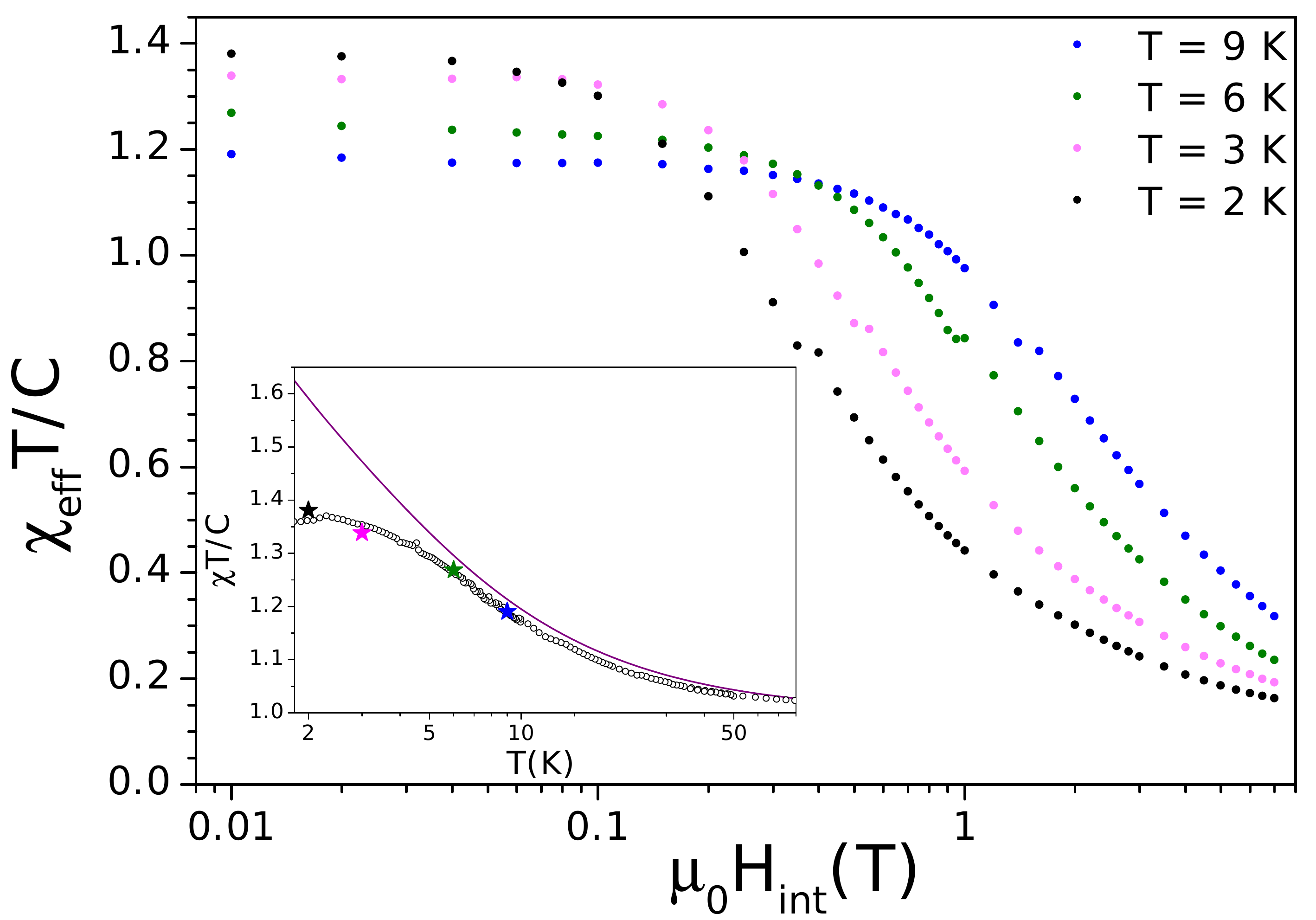}
\caption{The ratio $\chi_{\rm eff} T/C$ where $\chi_{\rm eff} = M/H$, as a function of field, showing that the susceptibility is only approximately linear in the range $\mu_0 H = 0 - 0.1$ T. 
The inset reproduces data from Ref. \cite{Laura} (black circles), with the coloured stars indicating the lowest field values from the main Figure (same colour scheme maintained throughout). The full line is the theoretical estimate for an idealised spin ice in the Husimi tree approximation of Jaubert {\it et al.} (see Refs. \cite{Jaubert-TSF,Laura}).} 
\end{center}
\end{figure}

\end{document}